\documentstyle[prl,aps,multicol]{revtex}
\input{epsf}

\begin{document}
\newcommand{\tbeta}{{\tilde \beta}}
\newcommand{\tsigma}{{\tilde \sigma}}

\title{Regular dendritic patterns induced by non-local time-periodic forcing}

\author{T. B\"orzs\"onyi$^{1,2}$, T. T\'oth-Katona$^1$, \'A. Buka$^1$,
and L. Gr\'an\'asy$^1$} 
\address{$^1$Research Institute for Solid State Physics and Optics,
  Hungarian Academy of Sciences,  H-1525 Budapest, P.O.B.49, Hungary\\
$^2$ Groupe de Physique des Solides, CNRS UMR 75-88, Universit\'es 
Paris VI et VII, Tour 23, 2 place Jussieu, 75251 Paris Cedex 05, France
} 
%\date{\today}
\maketitle
\begin{abstract}
The dynamic response of dendritic solidification to spatially 
homogeneous time-periodic forcing has been studied. 
Phase-field calculations performed in two dimensions (2D) 
and experiments on thin (quasi 2D) liquid crystal layers show 
that the frequency of dendritic side-branching can be tuned by 
oscillatory pressure or heating. The sensitivity 
of this phenomenon to the relevant parameters, the frequency 
and amplitude of the modulation, the initial undercooling and 
the anisotropies of the interfacial free energy and molecule 
attachment kinetics, has been explored. It has been demonstrated 
that besides the side-branching mode synchronous with external 
forcing as emerging from the linear Wentzel-Kramers-Brillouin 
analysis, modes that oscillate with higher harmonic frequencies 
are also present with perceptible amplitudes. 
\end{abstract}

%\twocolumn
\begin{multicols}{2}
\narrowtext
\section{Introduction}
\label{intro}

Complex patterns observed in nature attract a considerable 
interest recently \cite{hoh93,benjac94,goll99}. 
The complicated spatiotemporal behavior that leads to the formation 
of such patterns is usually associated with the instabilities of systems 
transforming under non-equilibrium conditions. A spectacular example of
practical importance is dendritic growth that appears in 
anisotropic systems where the propagation of the transformation 
front is coupled with the diffusion of a conserved quantity. 
The respective diffusional instabilities lead to the formation of a 
traveling quasi-steady-state tip that emits surface undulations which 
evolve into side arms in directions determined by the anisotropy of 
the system \cite{hu93}. Many of the technologically important 
materials form by dendritic solidification \cite{hu93}. Analogous 
phenomena have been reported in biological systems \cite{jash96}, 
anisotropic Hele-Shaw cells \cite{jaga89}, and in cosmology/particle 
physics \cite{neme90,adam93}. Although experiments on the freezing of 
transparent liquids \cite{ice,hugl81,glsi89,rugl91,bibi95,lq} 
clarified many of the essential features of dendrite formation, 
important questions remained open. For example, the role played by thermal 
fluctuations in side-branching is the subject of continuing investigations 
\cite{hu93,bibi95,pila86,doka87,gepo98,kara99}. It is 
anticipated that a selective amplification of thermal noise 
is responsible for the side-branching and for the observed 
irregularity of dendritic patterns \cite{hugl81,glsi89,rugl91,glma1}. 

While the steady-state behavior of dendritic growth 
is understood fairly well in the framework of recent 
theories (microscopic solvability \cite{hu93,ampo87,la87-2,la87-1} 
and phase-field theories 
\cite{hu93,pefi90,koba93,wase93,koba94,kara96,kara98}), 
less is known of the dynamic response 
of the dendritic morphology to time-dependent external perturbations. 
The non-linear systems often develop regular patterns under periodic
forcing \cite{goll99}. Considering the inherent non-linearity 
of the equations describing dendritic growth, it is reasonable to 
expect that such periodic perturbations lead to resonance patterns 
that might be used to influence the growth morphology. Besides 
its scientific interest, a detailed understanding of such phenomena 
could open novel routes for designing materials for specific applications.

Time-periodic forcing of dendritic side-branching has already been 
realized via heating locally the dendrite tip by laser beam 
\cite{qicu90,wimu93,muwh95}, or by exposing it to an oscillatory flow 
field \cite{boch90}. Both methods lead to the formation of fairly regular 
morphologies in a suitable frequency range. However, these methods 
cannot be easily used for spatially homogeneous controlling of growth 
morphologies in large volumes. To circumvent this problem, we 
demonstrated \cite{boto99} the possibility for tuning the dendritic 
morphology by spatially homogeneous time periodic forcing via modulated 
pressure \cite{szabo} and heating, that change the undercooling 
instantaneously and homogeneously in large volumes. 

In this paper we present a detailed analysis of the dynamic response of 
dendritic solidification to oscillating pressure and heating. In identifying 
the resonance conditions and the most important process parameters, we 
rely on phase-field modeling, one of the most potent methods for 
describing morphology evolution. The theoretical predictions will be 
confronted with experiments on quasi-2D liquid crystal layers, 
known as suitable model materials \cite{lq,buto95,tobo96,bobu98}. 
The rest of this paper is structured as follows. In section II we briefly 
summarize the phase-field equations that incorporate external forcing, 
and introduce quantities for characterizing growth morphology. 
Section III is devoted to the experimental setup and the details of the 
measurement techniques. In section IV, we present the phase-field
predictions and compare them with the experimental results. In section
V, our findings are discussed in the light of theory and experiment on 
tip-localized forcing. Finally, a few concluding remarks 
will be made in section VI.

\section{Phase-field theory with external forcing}
\label{equations}

We investigate the growth of the crystalline phase into an undercooled
single component liquid at a temperature between the melting point and the 
hypercooling limit (``unit undercooling'') \cite{hyper}. In this temperature 
range, the thermal diffusion controls the growth process, as opposed to the 
(molecule attachment) kinetics controlled mechanism taking place below the 
hypercooling temperature \cite{hyper1}. Owing to the heat release during 
solidification, the crystalline phase heats up to its melting point, 
while the rest of the latent heat is transported into the cold liquid. 
Under such conditions, the planar front is inherently unstable against thermal 
perturbations despite the stabilizing effect of the interfacial free energy 
\cite{instab}.

The phase-field theory is a powerful tool to study moving boundary 
problems. It represents a dynamic extension of the Cahn-Hilliard 
theory of first-order phase transformations, in which the evolution 
of a non-conserved order-parameter is coupled to thermal or chemical 
diffusion \cite{hu93,pefi90,koba93,wase93,koba94,kara96,kara98}. 
Here we use a thermodynamically consistent version \cite{pefi90} 
(Model I of \cite{wase93}) which ensures locally positive entropy 
production and was modified to incorporate anisotropies of the 
interfacial free energy and the kinetic coefficient 
\cite{mcwh93}. The local state of the matter is represented by the 
phase-field $\phi (${\bf r}$,t)$ \cite{shox96}. 

To incorporate external forcing, the equations \cite{wase93,mcwh93} that 
describe the evolution of the phase-field and the temperature field 
$T (${\bf r}$,t)$  in 2D

\begin{eqnarray}
\label{phasefield1}
\tau(\theta) \frac{\partial \phi}{\partial t}=
Q(T)p'(\phi) - G'(\phi) - {\delta \over \delta \phi} 
\biggl \lbrace 
{1 \over 2} [\kappa(\theta)]^2 |\nabla \phi|^2 \biggl \rbrace,
\end{eqnarray}

\begin{equation}
\label{heatdiff1}
\lbrace c_p + [p(\phi)-1] L'(T) \rbrace \frac{\partial T}{\partial t}
+ L(T) p'(\phi) \frac{\partial \phi}{\partial t}=
k \nabla^2 T,
\end{equation}

\noindent have to be modified. 
Here $\tau$ is an empirical relaxation coefficient, whose inverse 
is an intrinsic interfacial mobility, and
$\delta \over \delta \phi$ stands for variation with respect to $\phi$.
Other notations are as follows:  
$p(\phi) = \phi^3 (10 - 15 \phi + 6 \phi^2)$, 
$\kappa^2/2$ the coefficient of the square-gradient
term in the entropy functional, 
that depends on the orientation represented by the angle $\theta$
(defined as $tan(\theta) = \partial_y \phi / \partial_x \phi$),
$G(\phi) = \phi^2 (1-\phi)^2 / 4e$, $e$ is a constant that determines 
the height of the intermediate maximum of the double well-potential, 
$Q(T) = \int_{T_m}^{T}{\frac{L(\eta)}{\eta^2}d\eta} = 
-\frac{\Delta G(T)}{T}$, $\Delta G(T)$ the Gibbs free energy 
difference between the liquid and solid, while $c_p$, $L(T)$, 
and $k$ are the specific heat of the liquid, the enthalpy 
difference between the liquid and the solid, and the thermal 
conductivity, respectively. The primed quantities denote 
derivatives with respect to the arguments.
Note that $\phi$ and  $p$ are dimensionless. The units of $k$ and
$\kappa^2$ are W m$^{-1}$ K$^{-1}$ and J m$^{-1}$ K$^{-1}$,
respectively; while $Q, G, c_p$ and $1/e$ are measured in
J m$^{-3}$ K$^{-1}$.

The dominant effect of the pressure modulation 
$p(t) = p_0 + \Delta p(t)$ 
is a variation of the melting point described by 
the Clausius-Clapeyron law 
$T_m(p) = T_{m}(p_{0}) + \Delta p \Delta V/S_{m}$, where $\Delta V$ is 
the volume change upon solidification, and $S_m$ is the entropy of
fusion. This enters into Eq. (1) via the Gibbs free energy 
difference which may be approximated 
as $\Delta G(T,p) \approx S_m [T_m(p)-T]$ 
at small undercoolings ($T_m - T \ll T_m$).
The other relevant quantities [specific heat, heat of 
fusion, and entropy of fusion $L'(T)$] are far less sensitive 
to the variation of pressure. For example, integrating the 
Clausius-Clapeyron relationship one obtains 
$\Delta L = \Delta p \Delta V [ 1 + \Delta c_p /S_m + 
T (\alpha_l V_l - \alpha_s V_s)/\Delta V]$,
where $\Delta c_p = c_{p,l} - c_{p,s}$ is the variation 
of the specific heat upon melting,
$\alpha_{l,s}$ are the thermal expansion coefficients, while 
subscripts $l$ and $s$ refer to the liquid and the solid respectively. 
Under typical conditions of our experiments [pressure amplitude
$\sim$ 2 bar, undercooling of 1 K, and physical properties of CCH3
(Appendix A)], the undercooling varies 
by $\sim 6 \%$, while the relative change of $L$ (and $S_m$) is only 
$\sim 10^{-3}$. We found that such pressure effects on $L$ and 
$S_m$ do not influence the results perceptibly; 
i.e. pressure modulation needs to be taken into account only 
via $Q(T)$ in Eq. (1). Modulated heating, in turn, appears 
as a time dependent source term in Eq. (2).

At this stage it is advantageous to introduce dimensionless variables. 
This is done similarly to that in \cite{wase93}. 
However, to avoid complications emerging from an oscillating melting 
point, the reduced temperature has been redefined in terms of a constant 
reference temperature $T_r$ 
as $u (${\bf r}$,t) = (T-T_r)/(T_r-T_{\infty})$, where $T_{\infty}$
is the initial temperature of the undercooled liquid. The mathematical 
problem is invariant to the choice of $T_r$ so far as 
$A = u_m = (T_m - T_r)/(T_r - T_{\infty})$ is inserted as 
in Eq. (3). (With $T_r=T_m$, the usual form of the phase-field 
model is recovered.) 

The relevant physical properties are combined into 
dimensionless parameters $\Delta = c_{p}(T_r - T_{\infty})/L$, 
$\alpha = \sqrt{2}\omega S_{m} L /(12c_{p}\sigma_o)$, 
$\tau_o = S_{m} D \beta_o/\sigma_o$, and $\epsilon = \delta/\omega$,
$\tilde\sigma^{'} ={\rm d}\tilde\sigma / 
{\rm d}\theta $, where $D$ is the thermal diffusion coefficient, 
$\delta$ the interface thickness, and $\omega$ a reference length 
comparable to the size of a well developed crystallite.
The anisotropies of the interfacial free energy $\sigma$ and 
the kinetic coefficient $\beta$ are represented by the dimensionless 
functions $\tsigma = 1 + \sigma_{n}cos(n\theta)/2$ and 
$\tbeta = 1 + \beta_{n}cos(n\theta)/2$ of $n$-fold 
symmetry, that are related to the dimensional quantities via  
relationships $\sigma(\theta)=\sigma_o\tsigma(\theta)$ and 
$\beta(\theta)=\beta_o\tbeta(\theta)$. Here $\sigma_{n}$ and 
$\beta_{n}$ are the anisotropy parameters. Lengths and time are
scaled by $\omega$ and $\omega ^2/D$, respectively. The 
coefficient $\kappa$ and the constant $e$ are related 
to $\delta$ and the interfacial free energy \cite{wase93}. 

With these notations Eqs. (1) and (2) transform to

\begin{eqnarray}
\label{phasefield}
{\epsilon^2 \tau_o \tilde\beta  \tilde\sigma} 
\frac{\partial \phi}{\partial t} =
 \phi (1 - \phi)
\{\phi-{1 \over 2}+30 \epsilon \alpha \Delta [u-A(t)] 
\phi(1-\phi) \}
\nonumber \\*
-\epsilon^2 \frac{\partial}{\partial
x}[\tilde\sigma \tilde\sigma^{'} \frac{\partial 
\phi}{\partial y}]
             +\epsilon^2 \frac{\partial}{\partial
y}[\tilde\sigma \tilde\sigma^{'} \frac{\partial \phi}{\partial x}]
\nonumber \\*
             +\epsilon^{2} \nabla [\tilde\sigma^2  
\nabla \phi]
\end{eqnarray}
%\begin{multicols}{2}

\begin{equation}
\label{heatdiff}
{\frac{\partial u}{\partial t}}+{1 \over \Delta} 30 \phi^{2} 
(1-\phi)^{2}
\frac{\partial \phi}{\partial t} =  \nabla^{2} u + B(t),
\end{equation}

\normalsize
\noindent where the modulated pressure and heating are incorporated via 
terms $A(t)$ and $B(t)$, respectively. 

In this work Eqs. (\ref{phasefield}) - (\ref{heatdiff}) 
have been solved numerically on an $N \times N$ rectangular grid, 
$N = 1000$, that corresponds to an area $l \times l$ of dimensionless 
linear size $l = 5$ and a grid spacing $\Delta x=0.005$. An explicit 
finite difference scheme has been employed in the case of 
Eq. (\ref{phasefield}), while Eq. (\ref{heatdiff}) has been 
handled by the alternating-direction implicit method which is 
unconditionally stable \cite{mcwh93}. At $t=0$ a crystalline nucleus 
($\phi = 0$) is placed at the center of the model area filled by 
uniformly undercooled liquid ($\phi = 1$, $u = -1$). The reduced 
temperature and the phase-field were kept constant ($u = -1$ and 
$\phi = 1$) at the boundaries of the model space. To save CPU time, 
we used a primitive adaptive mesh technique: far from the solidification 
front Eq. (\ref{heatdiff}) has been solved on a rough grid (of spacing 
$10 \cdot \Delta x$). 
To model the effect of thermal fluctuations and 
to suppress undesirable lattice effects that favor specific growth 
directions, a spatially 
and temporally uncorrelated noise of amplitude
0.01 and zero mean value
has been added to the dimensionless temperature $u$ in every time step. 
If not stated otherwise, we use the dimensionless parameters: 
$\alpha = 350$, $\tau_o = 20$, 
$\epsilon = 0.005$, $\Delta x = 0.005$, and time step 
$\Delta t = 10^{-4}$. 
Owing to the known limitations of phase-field
modeling \cite{hu93,kipr99}, we performed our calculations at 
relatively large undercoolings, $\Delta = 0.40 - 0.68$.

To characterize the spatiotemporal behavior of the evolving dendritic 
morphology the following quantities were determined:

{\bf 1.} We measured the width $w$ of the dendrite behind the tip at a 
distance of $\zeta = 0.75$ (150 pixels). To investigate this feature 
for longer times, we performed separate simulations on a $1200 \times 600$ 
grid, oriented so that the larger dimension be parallel with the direction 
of growth (here we used $\Delta t = 8 \cdot 10^{-5}$). 
The amplitude $A_{\nu}$ in the Fourier transform $W(f)$ of $w(t)$ 
characterizes the response synchronous with external forcing. 

{\bf 2.} The symmetry of the growth patterns was characterized by
the quantities {\it a} and {\it b}: 
\begin{equation}
%\label{symm}
 a = \frac{\sum_{i,j=-N/2}^{N/2} 
(\phi_{i,j} - \phi_{-i,j})^2}{M}
\end{equation}
\begin{equation}
%\label{symm2} 
 b = \frac{\sum_{i,j=-N/2}^{0}
(\phi_{i,j} - \phi_{j,i})^2}{K}
\end{equation}
\noindent where $M$ and $K$ are the numbers of pixels in which
$0.4 < \phi < 0.6$ for the whole system and for the lower left quarter,
respectively. Accordingly, $a \ge 0$ measures the symmetry of the 
whole domain with respect to the $y-$axis, while $b \ge 0$ quantifies 
the axisymmetry of a main branch. Both parameters are zero for 
symmetric patterns.

\section{Experimental setup and conditions}

The nematic-smectic B ($N-S_B$) phase transition of liquid crystals 
is recognized as an appropriate model of crystallization 
in liquids \cite{buto95,tobo96,bobu98}. A specialty of this phase 
transformation is a large anisotropy of the interfacial free energy  
\cite{tobo96}, a property expected to be advantageous for the 
regularization of dendritic morphology on theoretical grounds
(see section IV.A and \cite{boto99}).

To test the predictions, experiments have been performed on thin 
CCH3 liquid crystal layers. Ready-made cells of E.H.C. Co. (Japan) 
KSRP-10 (of thickness 10 $\mu$m) and KSRP-02 (2 $\mu$m) 
have been filled with CCH3 (Merck, Darmstadt). The surface 
treatment of the bounding glass plates assured the planar 
alignment of both $N$ and $S_B$ phases [the directors {\bf n}$(N)$ 
and {\bf n}$(S_B)$ that describe the average orientation of the 
elongated molecules are in the plane of the cell], and the conducting 
layers on the bounding plates were used as electrodes. 

For pressure modulation the liquid crystal cell was placed into 
a brass box (see Fig. \ref{expfig1}) surrounded by a temperature controlled 
hot-stage of  accuracy $\pm$ 3 mK. The gas pressure in the brass 
box has been regulated by a computer controlled solenoid valve system that 
switches on and off an excess pressure $p_{\rm e}$ preset between 
0 to 2 bar with an accuracy of $\pm$ 0.03 bar. This allows square 
wave-like pressure modulations in the frequency range up to 
${\tilde \nu } \sim 2$ Hz. 

The modulated heat release in the bulk has been realized by 
periodically transmitting a high frequency (600 kHz) electric 
current through the liquid crystal layer produced by gating 
the signal of a function generator with ${\tilde \nu } \sim 1$ Hz. 
The local off--plane heat transport (a precondition of regulation 
with oscillatory heating, see section IV.A and \cite{boto99}) is 
ensured by the quasi-2D sample geometry and by the heat transport 
through the bounding glass plates. 

The growth patterns were monitored in transmitting mode via a 
polarizing microscope equipped with a CCD camera, a method relying 
on the different optical properties of the nematic and smectic 
B phases. The images were stored and processed by a PC. The spatial 
and time resolutions of the system were $512 \times 512$ pixels and 
0.04 s, respectively. The calibration procedure with 6.3 $\times$ 
objective and 3.2 $\times$ projector combination gave scale 
factors of $1.35 \pm 0.01$ $\mu$m/pixel in $x$ direction and 
$0.95 \pm 0.01$ $\mu$m/pixel in $y$ direction.

The same setup has been used to measure the pressure coefficient of
the temperature of transformation between the smectic-B and nematic 
phases (see Appendix A).

\section{Results}

\subsection{Theoretical predictions}
\label{simresults}

To find resonance patterns we investigated the parameter space defined 
by the frequency and amplitude of the modulation, the undercooling 
and the anisotropies of the interfacial free energy and the kinetic 
coefficient. The calculations were performed with four- and six-fold 
anisotropies ($n = 4$ and $6$ in expressions for $\tsigma$ and $\tbeta$).

Before presenting our results on non-local periodic forcing
we characterize the unmodulated {\it reference} state.

\subsubsection{Free dendritic growth} Independently of the symmetry of 
the system, a weak uncorrelated (noise-induced) side-branching occurs 
in the unmodulated reference states shown in Figs. \ref{simfig2}(a) 
and \ref{simfig3}(a). The width $w(t)$ of a main dendrite arm 
measured at a dimensionless distance of $\zeta = 0.75$ behind 
the tip [Fig. \ref{simfig4}(a)] indicates surface undulations 
that can be decomposed into a spectrum of oscillations which 
covers the dimensionless frequency range of $\sim 30$ to $\sim 
170$, centered around a characteristic frequency of $f_0 \sim 
100$ [see insert in Fig. \ref{simfig4}(b)]. This behavior accords 
with that seen in the phase-field simulations of Karma and 
Rappel \cite{kara99}, and with experimental results 
\cite{doka87,wimu93}.  In agreement with \cite{kara99}, we find 
that far from the tip the characteristic wave length 
$\lambda_0 = v/f_0$ depends much more weakly on 
the distance from the tip than predicted by the microscopic 
solvability theory \cite{la87-1,baba87}, a behavior interpreted 
in terms of the ``stretching'' of the perturbations \cite{kara99} 
that travel along the curved perimeter of the dendrite.

Considering that the applied uncorrelated noise probes the 
dynamic response of the system to a broad variety of 
frequencies, it is reasonable to seek ``resonance'' 
in the vicinity of the characteristic frequency 
$f_0$ of spontaneous side branching.

\subsubsection{Pressure modulations} 
According to the Clausius-Clapeyron law, the modulated pressure 
translates into a time dependent melting point, and is represented 
by inserting $A(t) = u_{m}(t) = a_{0}g(t)$ and $B(t) = 0$ 
into Eqs. (3) and (4). Since our experimental setup allows only 
square-wave modulations (characterized by the filling coefficient 
$\xi = t_{\rm on}/t_{0}$, where $t_{0}$ is the period of 
oscillations and $t_{\rm on}$ the pulse length), the simulations 
have been performed for this wave form (unless stated otherwise). 

(a) {\it The effect of forcing frequency:}
The frequency dependence of the growth patterns is shown in 
Figs. \ref{simfig2}(b)--(f) and \ref{simfig3}(b)--(f). 
At low forcing frequencies ($\nu < 15$), the system alternates 
between two steady-state growth modes yielding uncorrelated 
side-branching that correspond to the actual undercooling. 
It is, however, remarkable that the switching transient between 
the constant pressure stages initiates more pronounced 
side arms than formed during the constant pressure periods. 
Regular growth morphologies [Figs. \ref{simfig2}(b)--(e) 
and \ref{simfig3}(b)--(e)] are observed in the lower half  of dimensionless 
frequency range of spontaneous 
thermal side branching shown in the insert of Fig. \ref{simfig4}(b)]. 
At frequencies larger than these, the formation of side branches 
cannot follow the external forcing, and the uncorrelated thermal 
side-branching reappears [Figs. \ref{simfig2}(f) and \ref{simfig3}(f)]. 

In agreement with our earlier results \cite{boto99}, when regular 
morphologies are formed, the side-branching and the tip velocity 
correlate with the pressure modulation.

The parameters $w(t)$, $a$, and $b$ reflect the formation of 
regular morphologies. For example, a periodic variation of the 
width $w(t)$ of the dendrite is seen [Fig. \ref{simfig4}(a)], 
which correlates with the external sinusoidal forcing and leads 
to the formation of regular side branches. Note that besides the 
forcing frequency (that appears with a far larger amplitude in 
$W(f)$ than the spontaneous undulations), its second and third 
harmonics ($2\nu, 3\nu$) are also present albeit with an amplitude 
that diminishes for higher order harmonics [Fig. \ref{simfig4}(b)]. 

Apparently, periodic forcing with fixed frequency excites several 
surface modulations; one synchronous with the forcing frequency 
($\nu$) and others oscillating with the higher harmonic frequencies 
($2\nu, 3\nu,$ ...), which indicates a dynamic coupling among 
these modes associated with the non-linearity of the governing 
equations. 

The relative amplitudes of higher harmonics vary with the wave 
form of the pressure modulation yielding different growth patterns 
(see Fig. \ref{simfil}). Fairly similar patterns [{\it cf.} Figs. 
\ref{simfil} (a) and (b)] are observed for sinusoidal 
and square-wave modulations, provided that for the latter 
$\xi = 0.5$ [i.e., frequencies $(2n+1)\nu$, $n = 1,2,...$ are 
present in the forcing spectrum]. When the filling coefficient 
$\xi$ deviates from 0.5, the frequency/phase content of forcing 
varies (e.g., the second harmonic $2\nu$ appears) leading to such 
effects as the formation of side-branches of twice the number 
corresponding to the base frequency [{\it cf.} Fig. 
\ref{simfil}(a)--(c)]. To understand this behavior we analyze 
the frequency dependence of the dynamic response of the system 
to ``asymmetric'' forcing ($\xi = 0.3$).

The Fourier amplitudes $A_{\nu}$ and $A_{2\nu}$ (corresponding to 
modes that oscillate with the forcing frequency and its double, 
respectively) and the symmetry parameters $a$ and $b$ are presented
in Fig. \ref{simfig5} as a function of the forcing frequency $\nu$. 
$A_{\nu}$ and $A_{2\nu}$ show maxima if $\nu$ or $2\nu$ are
close to $f_{exc} \sim 70$ [Fig. \ref{simfig5}(a)], where $f_{exc}$ is
the frequency where the synchronous response has the maximum
amplitude. Note that $f_{exc}$ is somewhat below the characteristic
frequency $f_0 \sim 100$ of the unperturbed dendritic growth. A
possible explanation of this deviation might be a different 
``stretching'' of perturbations, related to differences seen between 
shapes of dendrite tips formed in free growth and under external 
forcing. It is also remarkable, that at low
forcing frequencies the amplitude $A_{2\nu}$ of the second harmonic 
becomes larger than $A_{\nu}$. This inversion of the relative
magnitudes of $A_{\nu}$ and $A_{2\nu}$ is responsible for the 
side-branch ``doubling'' shown in Fig. \ref{simfil}. A similar 
phenomenon is seen in Fig. \ref{simfig3}(b), where the second
harmonic is the dominant mode as indicated by the striking similarity of 
patterns corresponding to $\nu = 20$ and 40 [{\it cf.} Figs. 
\ref{simfig3}(b) and (c)]. Note the lack of alternating variation 
of the length of side-branches that appears when the synchronous 
mode is also present with a significant amplitude [as in Fig. 
\ref{simfil}(c)]. 

We find that the amplitude $A_{\nu}$ increases approximately 
exponentially with the distance $\zeta$ from the tip of the dendrite.
In contrast, $A_{2\nu}$ shows a maximum as a function of $\zeta$
(Fig. \ref{simamp}).

Remarkably, the symmetry parameters $a$ and $b$ display minima at 
lower forcing frequencies [around $\nu \sim 20$; see Fig. 
\ref{simfig5}(b)] than the maximum of the Fourier-amplitude 
$A_{\nu}$. This difference in the positions of the extrema is a 
manifestation of the fact that axial symmetry and frequency content 
are different constituents of regularity. It appears that $A_{\nu}$ 
gives a closer representation of the regularity recognized by the 
human eye than the symmetry parameters. 

The overall frequency dependence of the pattern formation supports 
our anticipation that regularization via periodic forcing with fixed
frequencies is possible when the latter fall in the characteristic 
frequency range of noise-induced side branching. 

(b) {\it The effect of the amplitude of forcing:}
The Fourier amplitudes $A_{\nu}$ and $A_{2\nu}$ and the symmetry 
parameters show that with increasing forcing amplitude $a_0$ 
the regularity of the growth patterns increases (Fig. \ref{simfig6}). 
This is manifested in monotonously increasing $A_{\nu}$ and 
$A_{2\nu}$ that saturate for large $a_0$. At the same time, $a$ and $b$ 
decrease with $a_0$, reflecting the increasing symmetry 
of the patterns. 

(c) {\it The effect of undercooling:}
The response to a given pressure modulation correlates with the 
tendency for ``natural'' side-branching (Fig. \ref{simfig7}). 
While at large undercoolings (e.g., $\Delta = 0.65$), the formation 
of side-branches is rather intense, leading to well developed 
side-arms (both with or without forcing), at $\Delta = 0.5$ or 
below free growth produces essentially no side-branches, and 
external forcing is needed to trigger them, also reflected in 
their smaller size. Note the difference in spatial periodicity, 
that originates from the temperature dependence of the tip velocity. 

The variation of the Fourier amplitudes $A_{\nu}$ and $A_{2\nu}$ 
with undercooling ($\Delta$) is shown in Fig. \ref{simfig8}. 
The maximum of $A_{\nu}$ might be understood as follows. We found 
that with increasing undercooling $f_{exc}$ increases. Accordingly, 
a fixed forcing frequency $\nu$ produces the maximum amplitude at 
the undercooling for which $f_{exc} \sim \nu$. For the same reason, 
the maximum amplitude for the second harmonic $2\nu$ appears at a 
larger undercooling.

(d) {\it The effect of anisotropy:}
Theoretical considerations and numerical simulations show that 
anisotropy plays a central role in the formation of dendritic morphology.
It is, therefore, reasonable to expect that it has a similar importance in
the formation of regular patterns. This expectation is verified by the
dependence of growth forms on the magnitude of anisotropies in the system
(Fig. \ref{simfig9}). The growth patterns and variations of the 
Fourier amplitudes $A_{\nu}$ and $A_{2\nu}$, and of the symmetry 
parameters (Fig. \ref{simfig10}) indicate that the regularizing effect 
of external forcing decreases dramatically with vanishing anisotropy, 
i.e. without well defined orientational preferences the external 
perturbations are unable to drive the amoebae-like growth forms 
into a regular pattern. Note that the tip velocity decreases with 
decreasing anisotropies (note the difference of simulation times
the snapshots correspond to), the excitation with $\nu = 60$ 
generates periodic side-arms with increasing spacing. 

In these calculations we assumed that both the interfacial free energy and 
the kinetic coefficient are anisotropic. For the sake of simplicity, their 
ratio was usually kept constant $\beta_4/\sigma_4 = -2$. Our further 
investigations show, however, that resonance patterns form if 
{\it any} of these anisotropies (that of $\sigma$ or $\beta$) is 
sufficiently large.

\subsubsection{\it Modulated heating}

(a) {\it Alternating heating and cooling:} They lead to essentially 
the same type of resonance patterns as pressure oscillations 
(Fig. \ref{simfig11}), provided that that the net heat production
in a period is negligible. A remarkable difference is, however, that 
while the average tip velocity $v_0$ is essentially independent of the 
amplitude of pressure modulations, it decreases with the amplitude 
of heat production/extraction (Fig. \ref{simfig12}); a phenomenon
that might be associated with different efficiencies of heating and 
cooling.

(b) {\it Oscillatory heating:} The introduction of a local off--plane 
thermal transport described by $B(t) = b_{0} g(t) + h [u(${\bf r}$,t) 
- u_{\infty}]$ was necessary to prevent the melting of the crystal 
for the heating amplitudes needed to generate regular patterns. Here 
$h < 0$ is a dimensionless heat transfer coefficient that imitates 
heat transfer perpendicular to the plane of the liquid crystal 
layer towards the surrounding of reduced temperature $u_{\infty}$.
Since this term serves as a local heat sink, the in-plane thermal 
diffusion becomes less restrictive. This results in the formation 
of more compact objects (see Fig. \ref{simfig13}) with thick main 
arms and less developed side-branches. While with an increasing 
amplitude of the heat pulses the side-arms become more regular,
they become smaller due to the dissipated heat.

Much like pressure modulations, oscillatory heating leads to an
oscillating tip velocity. Owing to a decreasing average undercooling 
accompanied with increasing heating amplitude $b_0$, the period 
averaged velocity $v_0$ of the tip decreases roughly linearly 
with $b_0$ (Fig. \ref{simfig14}). At the same time, the amplitude 
of the oscillatory part of the velocity $v_{\nu}$ increases
about linearly.

In summary, our investigations imply that under well defined 
conditions both types of non-local forcing can be used to control 
dendritic growth.

\subsection{Experimental results}
\label{expresults}

In defining the experimental conditions, we utilized the 
results of the computer simulations. For CCH3, the time 
and spatial resolution of our experimental setup is optimal 
at an undercooling of $\sim$ 1 K. Comparable variations of 
the undercooling ($\sim$ 0.1 -- 0.2 K) are achievable by 
applying modulation amplitudes as high as 2 bar or 3 $\times 
10^{-4}$ W/cm$^2$, as estimated on the basis of the pressure 
coefficient of the equilibrium temperature for the 
nematic--smectic B phases (see Appendix A) and other 
relevant properties. 

The appropriate frequency range of forcing has been determined 
by Fourier analysis of the width $w(t)$ of a freely growing dendrite 
measured at 66.5 $\mu$m behind the tip. Without perturbation, 
the side-branching is essentially random [Figs. \ref{pfrek}(a) and 
\ref{elfrek}(a)]. The characteristic frequency, identified as 
the peak of the broad spectrum in the insert of Fig. \ref{expfouri},
is roughly 1.8 Hz. Accordingly, our experimental setup was designed 
to cover the frequency range of 0 -- 2 Hz.

\subsubsection{\it Pressure modulations} 
The experimental results for pattern formation in the 2 $\mu$m 
thick liquid crystal cell are summarized in Fig. \ref{pfrek}. 
Resonance patterns of fairly regular side-branches are observed 
[Figs. \ref{pfrek} (b)--(f)] at forcing frequencies comparable 
with the characteristic frequency of the unperturbed state 
$\tilde \nu \sim  1.8$ Hz. The figure shows germs that nucleated 
heterogeneously at the same site with essentially the same orientation 
several times, yielding reproducible patterns. Although the respective 
dimensionless undercooling ($\Delta = 0.06$) is considerably smaller 
than in the simulations ($\Delta = 0.4-0.68$), the observed behavior 
follows closely the predictions of the phase-field theory. This 
similarity is especially striking when patterns formed at similar 
reduced frequencies ($\tilde \nu /f_0$) are compared [{\it cf.} Figs. 
\ref{simfig2}(c) and \ref{pfrek}(b) of $\tilde \nu /f_0 = 0.40$ and 0.42; 
or Figs. \ref{simfig2}(d) and \ref{pfrek}(d) of $\tilde \nu/f_0 = 0.6$ 
and 0.65]. An interesting observation is that far from the dendrite tip 
irregularities develop in the lengths of the side-branches, despite 
the even distance of their trunks. We observed an analogous phenomenon 
driven by the interaction with neighboring side-arms via thermal 
diffusion fields in long-time phase-field simulations \cite{movie}.

The effect of the {\it forcing amplitude} is shown 
in Fig. \ref{expamp}. Pressure oscillations of amplitude of 1 bar 
efficiently regularize the side-branch formation. With increasing 
amplitude the phenomenon becomes more pronounced as predicted by 
the phase-field calculations. At the largest pressure amplitudes 
even the secondary side-branches correlate with the pressure 
oscillations [see Fig. \ref{expamp}(c)]. Note again that the 
competition of neighboring side-branches leads to irregularities
in their lengths. 

\subsubsection{Modulated heating} 
In analogy to pressure modulations, the periodic heating experiments 
on the 10 $\mu$m thick liquid crystal cell also reveal regular 
side-branching [see Figs. \ref{elfrek}(a)--(c)]. The formation of 
side-branches correlates with the external forcing as illustrated 
in Fig. \ref{elfrek}(b), where the black lines denote the position 
of the tip at the centers of the heating pulses. The correlation is 
also evident from the power spectrum of the width of the dendrite 
measured 66.5 $\mu$m behind the tip (Fig. \ref{expfouri}). In full 
accord with the phase-field simulations for asymmetric square
waves [Fig. \ref{simfig4}(b)], the power spectrum indicates the 
presence of modes that are either synchronous with the forcing 
frequency ${\tilde \nu}$ = 0.46 Hz, or oscillate with doubled 
frequency ($2 \tilde \nu$). Even a peak corresponding to the 
third harmonic ($3 \tilde \nu$) may be identified, although with 
an amplitude that is close to the experimental uncertainty. 

Remarkably, in the response to ``symmetric'' forcing $\xi = 0.5$, 
the $2 \tilde{\nu}$ mode is also present (see the Fourier-spectrum and the 
short side-arms of doubled frequency in Fig. \ref{expdouble}), 
although this frequency is absent from the forcing spectrum. 
This finding confirms the non-linear behavior revealed by 
phase-field simulations, that higher harmonics missing from the 
forcing spectrum are also excited [see $W(f)$ for sinusoidal 
forcing in Fig. \ref{simfig4}].

Increasing the {\it heating amplitude or $\xi$} so that 
the period-averaged heating power reaches 
${\overline P} \approx \rm 10^{-4}$ W/cm$^2$, 
the formation of the side-branches is suppressed 
[see Fig. \ref{elfrek}(d)], a phenomenon resembling
that seen in the numerical simulations (Fig. \ref{simfig13}). 
Further increase of the heating amplitude (and power) melts 
the dendrites back. 

In line with our theoretical predictions (see Fig. 2. of 
\cite{boto99}), a weak oscillation of the tip velocity has been 
observed that correlates with the forcing. However, its amplitude 
is just above the resolution of the present experimental setup. 

A remarkable feature of the "regularized" dendrites, not seen in the
simulations, is a shift in the position of the side-branches on the 
two sides of the main tip [see e.g. Figs. \ref{elfrek}(b) and (c)]. 
As a result, one cannot use the symmetry parameters $a$ and $b$ 
[defined by eqs. (3) and (4)] for characterization of the pattern's 
regularity. For the same reason, the Fourier spectra shown in Figs. 
\ref{expfouri} and \ref{expdouble} have been evaluated from the 
half-width of the dendrite. Despite the shift of the side-branches, 
the power spectra on the two sides are fairly similar.

The shift in the position of the side-branches might be attributed 
to the asymmetry of the dendrite tip (due to the angular 
dependence $\sigma(\theta)$ [35]), shown by a high resolution 
snapshot of the tip region of a dendrite grown freely at 
$\Delta T = 1.0$ $^{\circ}$C (Fig. \ref{tipshape}). Another remarkable 
feature is that the left side of the tip is faceted. The first 
observable surface undulation (marked with arrow in Fig. \ref{tipshape}) 
that evolves later into a side-arm appears on the opposite side. In 
contrast, the faceted side remains smooth up to the same distance 
behind the tip.

Finally, one should mention that the electric heating in liquid crystals 
may have side-effects that are not incorporated into our phase-field 
model. Switching the electric field on, the orientation of the 
nematic director {\bf n}$(N)$ changes from planar to homeotropic 
(perpendicular to the bounding plates), an effect that influences 
the magnitude and anisotropy of the interfacial free energy, 
and may induce local flow in the sample. However, these side-effects
are of minor importance, since they are present in full strength 
much below the electric field needed for regularization.

\section{Discussion}

In this section we confront our results with those on the dynamic 
response of dendritic growth to local forcing with fixed frequency
as emerging from the microscopic solvability theory (MST) and experiment. 
It is appropriate to mention, that some features of the MST 
derivation prevent a quantitative comparison. For example, the MST 
calculations were performed for a 2D symmetric dendrite, while the 
anisotropy, the stability coefficient, and the P\'eclet-number were 
assumed to be small; conditions that are not met in our simulations 
and experiments. Therefore, a qualitative comparison is only meaningful. 
An important further difference between previous work and ours is that 
we used non-local forcing, as opposed with the tip-localized forcing
assumed in the MST.

Let us first recall some of the MST predictions on noise amplification 
(based on the Wentzel-Kramers-Brillouin approximation) that have 
been tested by comparison with numerical simulations \cite{baba87,kele87} 
and experiments \cite{wimu93} for fixed frequency perturbations localized 
at the tip.

(i) Periodic forcing by a fixed frequency $\nu$ leads to 
surface oscillations of amplitude that increases exponentially 
with the distance from the dendrite tip up to a critical distance 
$\zeta_c$ proportional with $1/\nu^4$. For larger distances the 
amplitude decreases and eventually dies away. If, in turn, the 
distance from the tip is fixed and $\nu$ is varied, a peak is 
observed in the amplitude.

(ii) The localized wave packets behave differently; they grow 
exponentially as they move to arbitrarily large distances from 
the tip, while the respective characteristic wave length increases. 

Our simulations are in line with the previous MST and numerical 
predictions for tip-localized forcing in the following respects: 

(a) In the vicinity of the dendrite tip, the amplitude of the mode 
synchronous with forcing increases roughly exponentially with 
the distance $\zeta$ from the tip. Owing to the excessive 
computation time and memory needed, we were unable to study the 
decay of this mode in detail. Nevertheless, as expected from the MST,
at large frequencies [Fig. \ref{simfig2}(f) and \ref{simfig3}(f)] 
the synchronous mode can only be recognized in the vicinity of
the dendrite tip. Another sign, that accords with the presence of 
a critical distance $\zeta_c$ beyond which the excited mode decays, 
is a maximum observed in the amplitude $A_{2\nu}$ of the second 
harmonic (Fig. \ref{simamp}) in simulations where the forcing 
spectrum contains $2\nu$ with a significant amplitude.

(b) The amplitude $A_{\nu}$ measured at a fixed distance behind 
the dendrite tip shows a maximum as a function of the forcing 
frequency [Fig. \ref{simfig5}(a)]. 

An interesting feature, revealed by our phase-field simulations, is 
that besides the synchronous mode, the Fourier spectrum of the dendrite 
width contains the second and third harmonics with perceptible amplitudes. 
This finding is confirmed by our experiments. We believe that this is 
the first direct demonstration of such non-linear effects in connection with 
diffusional instability induced dendrite formation during a first-order 
phase transformation. It appears, that this non-linear behavior is not a 
peculiarity of spatially homogeneous forcing. Although not mentioned in 
the original works, traces of higher harmonics seem to be present in the 
respective power spectra for local forcing as well \cite{boch90,wimu93}.
It is worth mentioning, furthermore, that a similar behavior has been 
observed in the case of anomalous Saffman-Taylor fingering \cite{raco88}, 
suggesting that the appearance of higher harmonic modes in response to 
periodic external forcing is a common feature of dendrite formation 
whether driven by diffusional or mechanical instabilities.

Summarizing, in the present state of affairs it seems that the only 
specific feature of pattern formation under non-local forcing is a 
simultaneous triggering of side-branches (of first, second, and higher 
orders) on independent crystallites throughout the sample.

\section{Summary}
\label{concl}

Our computer simulations and experiments demonstrated that the 
dendritic morphology can be regularized by non-local time-periodic 
forcing realized by modulated pressure and Joule heating. These 
conditions lead to an oscillatory velocity of the dendrite tip, 
and yield side-branches at regular distances, provided that the 
frequency of the modulation is close to the ``natural'' frequency 
of free side-branch formation. The dynamic response of the system 
to such non-local forcing can be understood in general on the basis 
of previous theoretical and experimental results on tip-localized 
forcing. However, the non-linear effects, such as the appearance 
of higher harmonic modes, warrant further theoretical work.

\acknowledgments{
The authors express their thanks to Dr. Gy\"orgy Szab\'o (Research
Institute for Materials Science, Budapest) for calling their
attention to the possibility of regulating dendritic morphology
by pressure modulations, and to Dr. N\'andor \'Eber (Research Institute
for Solid State Physics and Optics, Budapest) for his help in 
solving technical problems. This work has been supported by research grants 
OTKA T025139, T031808, F022771 and EU HPMF-CT-1999-00132. 
Part of the computations were performed 
on computers donated by the Alexander von Humboldt Foundation.
}

\begin{appendix}

\section{Pressure dependence of the phase transition temperature} 

The pressure dependence of the $N - S_B$ phase transition temperature 
has been measured by a procedure similar to that described in 
\cite{tobo96}. For different $p_e$ a single $S_B$ monodomain has been 
kept at constant size by controlling the temperature for several hours 
(until the thermodynamic equilibrium state of the system has approached). 
Fig. \ref{append1} shows the change of the phase transition temperature 
depending on the pressure. Linear fit on the data gives a slope 
${{\rm d} T_m \over {\rm d} p} = (0.032 \pm 0.003)$ K/bar for the 
Clapeyron coefficient. 

The Clapeyron coefficient can be also calculated from the relation: 
\begin{equation}
{{\rm d} T_m \over {\rm d} p} = {T_m^0 \Delta V_{N \rightarrow SB} 
\over \Delta H} 
\end{equation}
where $T_m^0$, $\Delta V_{N \rightarrow SB}$ and 
$\Delta H$ are the phase transition temperature at atmospheric pressure, 
the molar volume change on transition and the molar latent heat of fusion, 
respectively. From the relevant material parameters of CCH3, namely $T_m^0 = 
329.45$ K \cite{tobo96}, $m = 233$ (molar mass), $\rho = 895.4$ kg/m$^3$ 
(density at 80.6 $^{\circ}$C \cite{asahi91}), $\Delta H = 6247$ J/mol, 
$\Delta V_{N \rightarrow SB}/V = -2.6$ $\%$, and $\alpha_N = 25 \times 
10^{-4}$ 1/K (volumetric expansion coefficient in the nematic phase 
\cite{grasso89}) one obtains ${{\rm d} T_m \over {\rm d} p} = 0.033$ 
K/bar in excellent agreement with the experimental value.

Although this value is much larger than for metals, it is not unusual for 
liquid crystals. For example, $\sim$ 0.03 K/bar has been reported for the 
nematic -- crystal transition in PAA, and $\sim$ 0.1 K/bar for the 
nematic smectic transition in p-methoxybenzoic acid \cite{shas75}.
Comparable, or even larger coefficients have been measured on other organic 
substances such as camphene 0.214 K/atm, pivalic acid 0.0674 K/atm, and 
succinonitrile 0.0245 K/atm \cite{coko98}.

\end{appendix}

%%%%%%%%%%%%%%%%%%%%%%%%%%%%%%%%%%%%%%%%%%%%%%%%%%%%%%%%%%%%%%%%%%%%%%

%\bibliographystyle{normal}
%\bibliography{pattern}

\begin{references}
\bibitem{hyper} Hypercooling limit/unit undercooling: that 
temperature/undercooling at which the enthalpy of freezing heats the solid 
up to its melting point, $T_{m} - T_{hyper} = L/c_p$.
\end{references}

\begin{minipage}{8.3cm}
\begin{figure*}
\epsfxsize=8cm
\centerline{\epsfbox{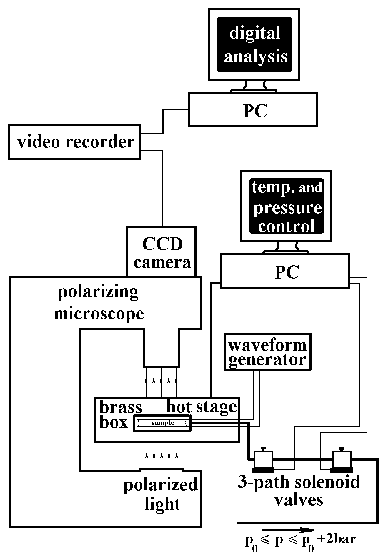}}
\caption{The experimental setup.
}
\label{expfig1}
\end{figure*}
\end{minipage}

\begin{minipage}{8.3cm}
\begin{figure*}
\epsfxsize=8cm
\centerline{\epsfbox{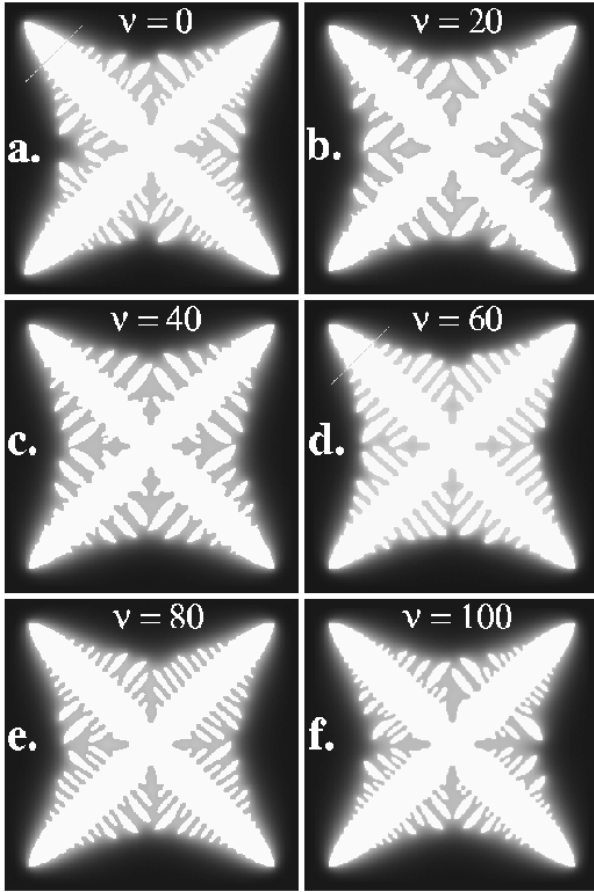}}
\caption{
Dendritic patterns of four-fold symmetry predicted by the phase-field
theory for $\Delta = 0.6$, $\sigma_4=-0.12$,  $\beta_4=0.24$ and 
$t = 0.22$: (a) Under constant pressure yielding $A = u_m = 0.048$; 
and (b)--(f) in the presence of square-wave pressure modulations of
amplitude $a_0 = 0.16$ and filling coefficient $\xi = 0.3$.  The gray 
level in the liquid phase indicates the temperature field. The white 
line marks the position, where the width of the dendrite arm was 
measured.
}
\label{simfig2}
\end{figure*}
\end{minipage}

\begin{minipage}{8.3cm}
\begin{figure*}
\epsfxsize=8cm
\centerline{\epsfbox{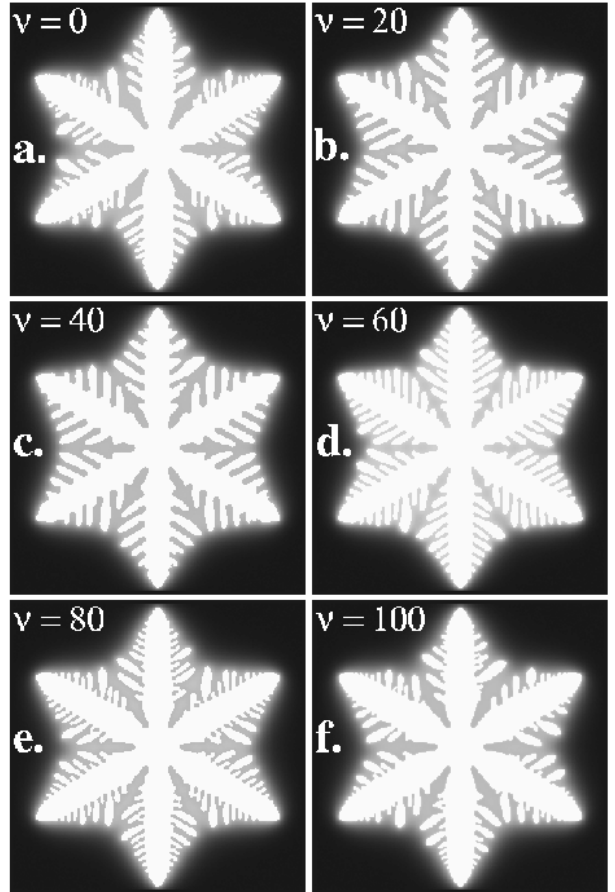}}
\caption{
Dendritic patterns of six-fold symmetry predicted by the phase-field
theory for $\Delta = 0.6$, $\sigma_6=-0.04$,  $\beta_6=0.08$ and 
$t = 0.22$:  (a) Under constant pressure yielding $A = 0.048$; 
and (b)--(f) in the presence of square-wave pressure modulations of
amplitude $a_0 = 0.16$ and filling coefficient $\xi = 0.3$. 
}
\label{simfig3}
\end{figure*}
\end{minipage}

\begin{minipage}{8.3cm}
\begin{figure*}
\epsfxsize=8cm
\centerline{\epsfbox{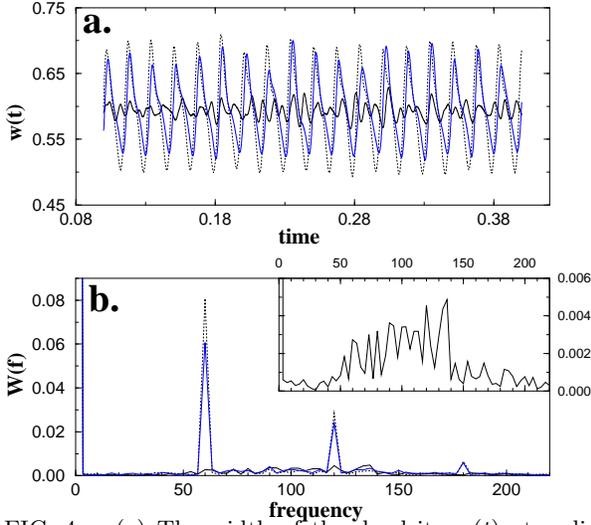}}
\caption{
(a) The width of the dendrite $w(t)$ at a distance of $\zeta = 0.75$ 
behind the dendrite tip without periodic perturbation [heavy solid 
line; Fig. \ref{simfig2}(a)], in the presence of sinusoidal (light 
solid line) and square-wave [dotted line; Fig. \ref{simfig2}(d)]. 
(b) The respective Fourier spectra $W(f)$. 
}
\label{simfig4}
\end{figure*}
\end{minipage}

\begin{minipage}{8.3cm}
\begin{figure*}
\epsfxsize=8cm
\centerline{\epsfbox{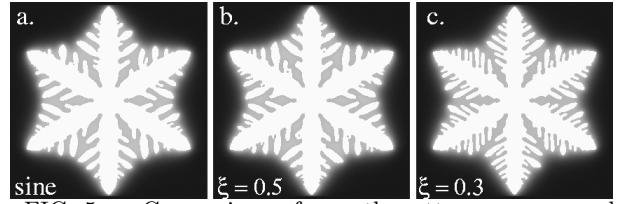}}
\caption{
Comparison of growth patterns corresponding to (a) sinusoidal, 
(b) symmetric ($\xi = 0.5$) and (c) asymmetric ($\xi = 0.3$) square-wave 
forcing of equal period lengths ($\nu = 30$), and period averaged 
amplitudes. Note the side-branch ``doubling'' in panel (c). (Other 
relevant parameters are as for Fig. \ref{simfig3}.) 
}
\label{simfil}
\end{figure*}
\end{minipage}

\begin{minipage}{8.3cm}
\begin{figure*}
\epsfxsize=8cm
\centerline{\epsfbox{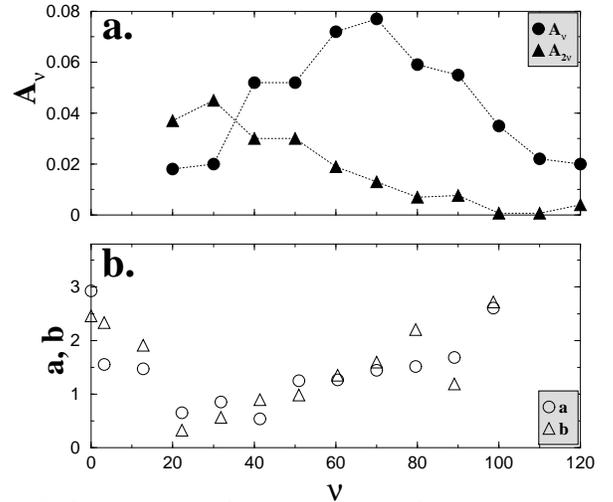}}
\caption{
The amplitude of the Fourier peaks $A_{\nu}$ and $A_{2\nu}$ 
measured at $\zeta = 0.75$ behind the tip, and the symmetry parameters 
$a$ and $b$ as a function of the driving frequency $\nu$. Other relevant 
parameters are $\Delta = 0.55$, $\xi = 0.3$, $\sigma_4 = -0.12$ and 
$\beta_4 = 0.24$, while $a_0 = 0.14$ (a), and $a_0 = 0.2$ (b).
}
\label{simfig5}
\end{figure*}
\end{minipage}

\begin{minipage}{8.3cm}
\begin{figure*}
\epsfxsize=8cm
\centerline{\epsfbox{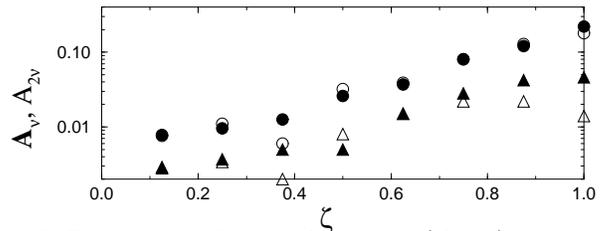}}
\caption{
The Fourier amplitudes $A_{\nu}$ (circles) and $A_{2\nu}$ (triangles) 
as a function of the distance $\zeta$ from the dendrite tip at two
undercoolings $\Delta = 0.55$ (full symbols) and 0.6 (empty symbols).
Note the roughly exponential behavior of $A_{\nu}$ and the 
saturation/maximum in $A_{2\nu}$. 
Other relevant parameters are 
$\nu = 60$, $\xi = 0.3$, $a_0 = 0.16$,  $\sigma_4 = -0.12$ and $\beta_4 = 0.24$.
}
\label{simamp}
\end{figure*}
\end{minipage}

\begin{minipage}{8.3cm}
\begin{figure*}
\epsfxsize=8cm
\centerline{\epsfbox{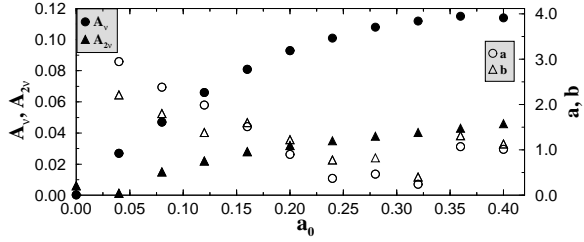}}
\caption{
The amplitude of the Fourier peaks $A_{\nu}$ and $A_{2\nu}$, and the 
symmetry parameters $a$ and $b$ as a function of the forcing amplitude 
$a_0$ for the system shown in Fig. \ref{simfig2}(d).
}
\label{simfig6}
\end{figure*}
\end{minipage}

\begin{minipage}{8.3cm}
\begin{figure*}
\epsfxsize=8cm
\centerline{\epsfbox{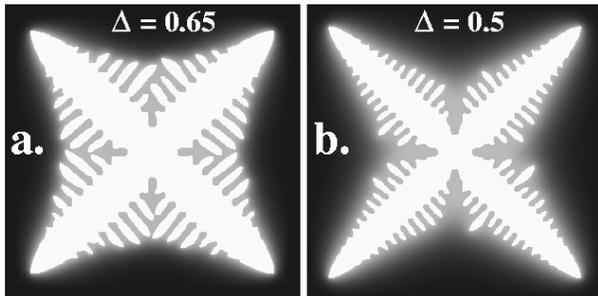}}
\caption{
The effect of undercooling on pattern formation under
periodic forcing. (a) $\Delta = 0.65$; (b) $\Delta = 0.5$.
Other parameters as for Fig. \ref{simfig2}(d) except 
that $t = 0.2$ and $t = 0.283$ for panels (a) and (b), 
respectively. 
}
\label{simfig7}
\end{figure*}
\end{minipage}

\begin{minipage}{8.3cm}
\begin{figure*}
\epsfxsize=8cm
\centerline{\epsfbox{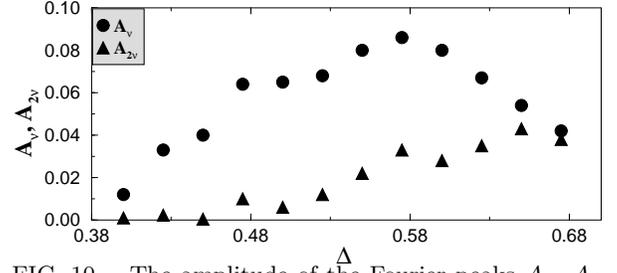}}
\caption{
The amplitude of the Fourier peaks $A_{\nu}$, $A_{2\nu}$ 
as a function of undercooling $\Delta$. 
Other relevant parameters are $\nu = 60$, $\xi = 0.3$, $a_0 = 0.16$, 
$\sigma_4 = -0.12$ and $\beta_4 = 0.24$.
}
\label{simfig8}
\end{figure*}
\end{minipage}

\begin{minipage}{8.3cm}
\begin{figure*}
\epsfxsize=8cm
\centerline{\epsfbox{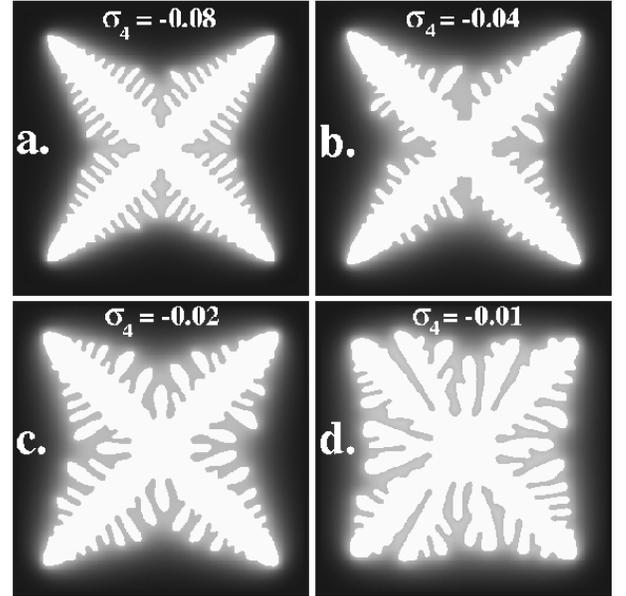}}
\caption{
Pattern formation under periodic forcing as a function of 
anisotropy: $\sigma_4=-0.08$ (a), $\sigma_4=-0.04$ (b), 
$\sigma_4=-0.02$ (c) and $\sigma_4=-0.01$ (d).
Other relevant parameters are 
$\Delta = 0.55$, $\nu = 60$, $\xi = 0.3$, $a_0 = 0.16$ 
and $\beta_4 = -2\sigma_4$.
The average growth rate varies with anisotropy, therefore,
different times are needed to reach comparable sizes: 
$t=0.24$ (a), $t=0.3$ (b), $t=0.38$ (c), and $t=0.44$ (d).
}
\label{simfig9}
\end{figure*}
\end{minipage}

\begin{minipage}{8.3cm}
\begin{figure*}
\epsfxsize=8cm
\centerline{\epsfbox{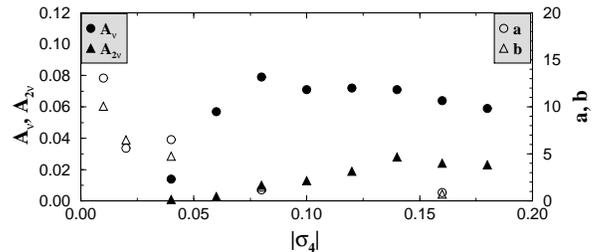}}
\caption{
The amplitude of the Fourier peaks $A_{\nu}$, $A_{2\nu}$,
and the symmetry parameters $a$ and $b$ as a function of 
anisotropy parameter $|\sigma_4|$ of the interfacial free 
energy. The kinetic anisotropy has been varied
proportionally ($\beta_4 = - 2\sigma_4$). 
Other relevant parameters are $\Delta = 0.55$, $\nu = 60$,  
$\xi = 0.3$ and $a_0 = 0.16$.
}
\label{simfig10}
\end{figure*}
\end{minipage}

\begin{minipage}{8.3cm}
\begin{figure*}
\epsfxsize=8cm
\centerline{\epsfbox{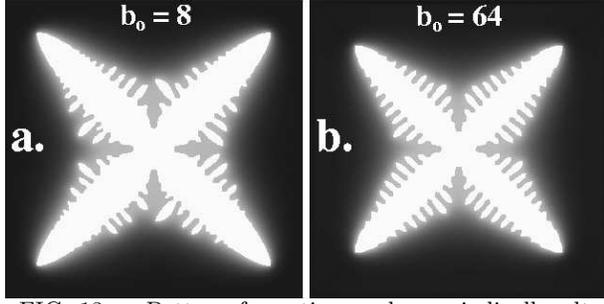}}
\caption{
Pattern formation under periodically alternating heating and cooling of 
amplitudes $b_0 = 8$ (a) and 64 (b). Other relevant parameters are 
$\Delta = 0.6$, $\nu = 60$, $\xi = 0.5$, $\sigma_4=-0.12$, $\beta_4=0.24$,
while the elapsed time $t$ = 0.22.
}
\label{simfig11}
\end{figure*}
\end{minipage}

\begin{minipage}{8.3cm}
\begin{figure*}
\epsfxsize=8cm
\centerline{\epsfbox{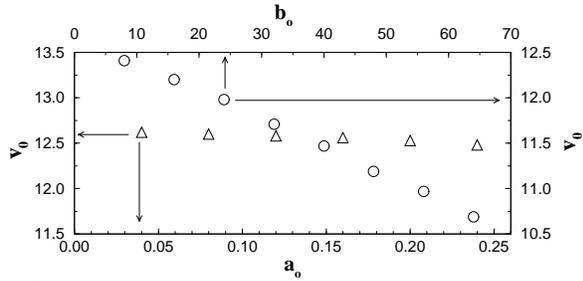}}
\caption{
The average velocity of the tip under periodic heating and 
cooling (open circles; parameters as for Fig. \ref{simfig11}), 
and for oscillatory pressure (open triangles; calculated with
$<p> = 0, \xi = 0.5,$ other parameters as for Fig. \ref{simfig2}(d)).
}
\label{simfig12}
\end{figure*}
\end{minipage}

\begin{minipage}{8.3cm}
\begin{figure*}
\epsfxsize=8cm
\centerline{\epsfbox{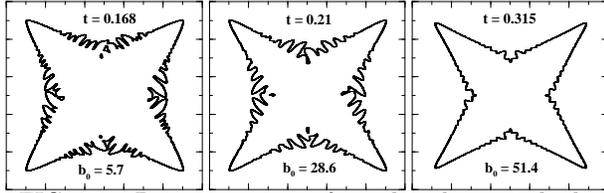}}
\caption{
Pattern contours formed under periodic heat pulses of amplitudes 
increasing from left to right ($b_0 = 5.7$, $b_0 = 28.6$, and 
$b_0 = 51.4$). Other relevant parameters are $\Delta = 0.61$, 
$\nu = 50$, $\xi=0.15$, $\sigma_4 = -0.16$, $\beta_4 = 0.32$, 
$h=-14.3$, $\tau_o = 16$ and $\Delta t = 7\times 10^{-5}$.  
}
\label{simfig13}
\end{figure*}
\end{minipage}

\begin{minipage}{8.3cm}
\begin{figure*}
\epsfxsize=8cm
\centerline{\epsfbox{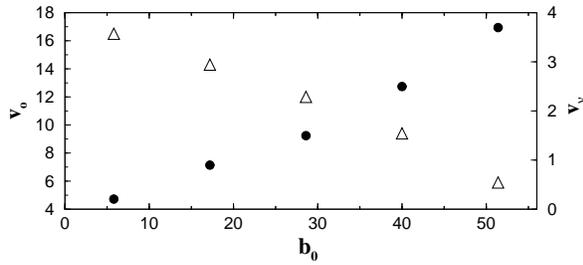}}
\caption{
The average velocity $v_o$ (open triangles) and the amplitude of 
velocity oscillation $v_{\nu}$ (full circles) as a function of 
the heating amplitude $b_0$ for simulations presented in 
Fig. \ref{simfig13}.
}
\label{simfig14}
\end{figure*}
\end{minipage}

%%%%%%%%%%%%%%%%%%%%%%%%%%%%%%%%%%%%%%%%%%%%%%

\begin{minipage}{8.3cm}
\begin{figure*}
\epsfxsize=8cm
\centerline{\epsfbox{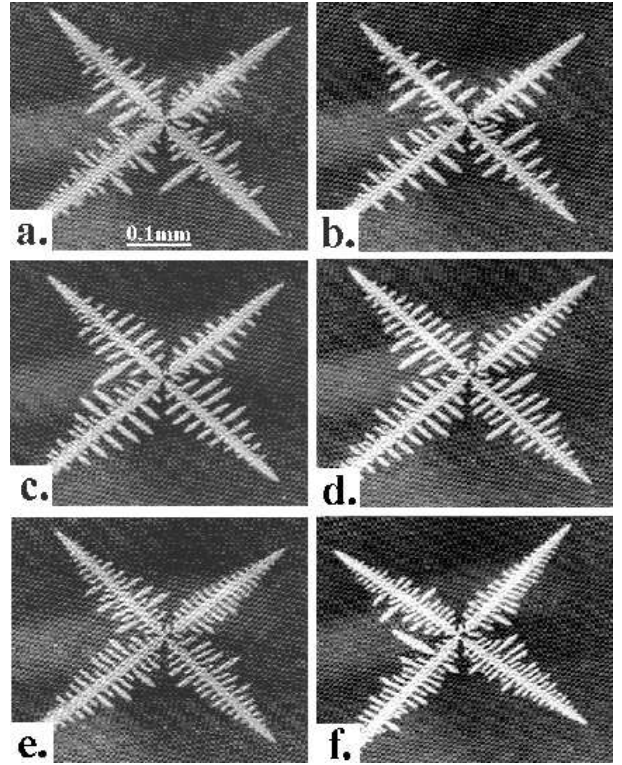}}
\caption{
The effect of oscillatory pressure on the smectic-B dendrite 
growing into undercooled nematic phase ($\Delta T$ = 1.0 $^{\circ}$C). 
(a) Without pressure oscillations; (b)--(f) with square wave 
pressure modulation of parameters $p_e$ = 2 bar, $\xi = 0.2$, 
${\tilde \nu} =$ 0.75, 0.91, 1.18, 1.64, and 1.96 Hz, 
respectively. 
}
\label{pfrek}
\end{figure*}
\end{minipage}

%\end{multicols}

%\begin{multicols}{2}
%\narrowtext

\begin{minipage}{8.3cm}
\begin{figure*}
\epsfxsize=8cm
\centerline{\epsfbox{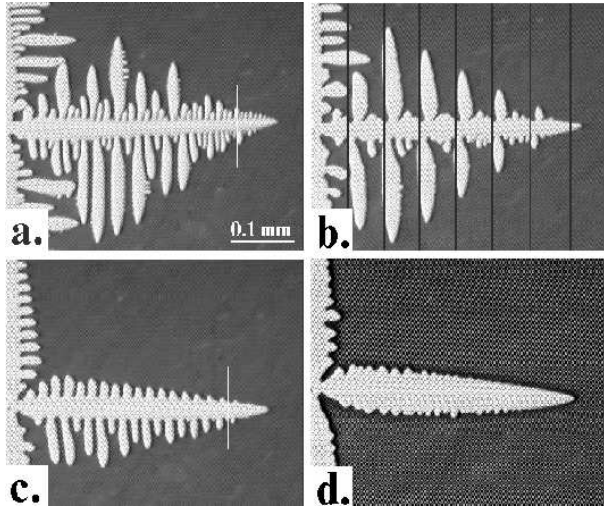}}
\caption{
The effect of the oscillatory heating on smectic-B dendrite 
growing into undercooled nematic phase ($\Delta T = 
1.0^{\circ}{\rm C}$). 
(a) No oscillatory heating; 
(b) ${\tilde \nu}$ = 0.2 Hz, 
${\overline P}= 6 \times 10^{-5}$ W/cm$^2$, $\xi = 0.14$; 
(c) ${\tilde \nu}$ = 0.46 Hz, 
${\overline P}= 6 \times 10^{-5}$ W/cm$^2$, $\xi = 0.16$; 
(d) ${\tilde \nu}$ = 0.81 Hz, 
${\overline P}= 1.1 \times 10^{-4}$ W/cm$^2$, $\xi = 0.61$. 
The two lines in panels (a) and (c) mark the positions, 
where the width $w$ of the dendrite has been measured as a 
function of time (for the Fourier transforms see Fig. \ref{expfouri}). 
Black lines in panel (b) denote the position of the tip 
at the centers of the heating pulses.  
}
\label{elfrek}
\end{figure*}
\end{minipage}

%\newpage

\begin{minipage}{8.3cm}
\begin{figure*}
\epsfxsize=8cm
\centerline{\epsfbox{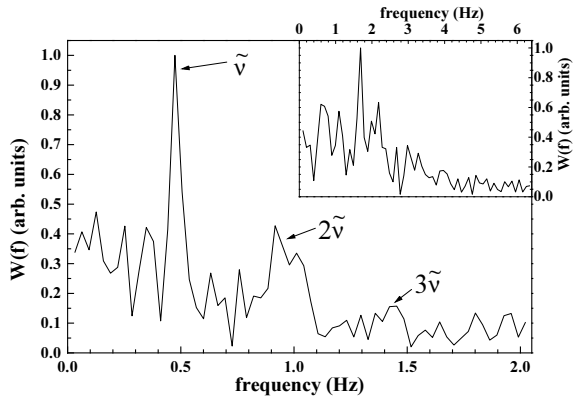}}
\caption{Fourier spectrum $W(f)$ of the half-width of the dendrite
measured 66.5 $\mu$m behind the tip as shown in Fig. \ref{elfrek}(c). For 
comparison, the respective spectrum for free dendritic growth 
[Fig. \ref{elfrek}(a)] is also presented (see insert). 
}
\label{expfouri}
\end{figure*}
\end{minipage}

%\newpage

%\begin{minipage}{16.6cm}
\begin{minipage}{8.3cm}
\begin{figure*}
\epsfxsize=8cm
\centerline{\epsfbox{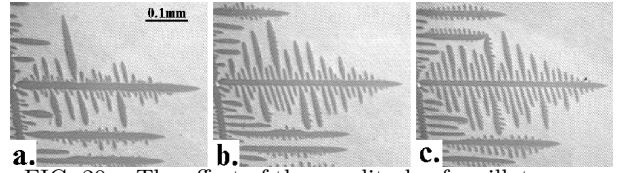}}
\caption{
The effect of the amplitude of oscillatory pressure on 
the smectic-B dendrite growing into undercooled nematic 
phase ($\Delta T$ = 1.0 $^{\circ}$C). (a) Without pressure 
oscillations; and with square wave pressure modulation 
of parameters ${\tilde \nu}$ = 1.64 Hz, $\xi = 0.2$, 
$p_e =$ 1 bar (b) and 2 bar (c), respectively. 
}
\label{expamp}
\end{figure*}
\end{minipage}

\begin{minipage}{8.3cm}
\begin{figure*}
\epsfxsize=8cm
\centerline{\epsfbox{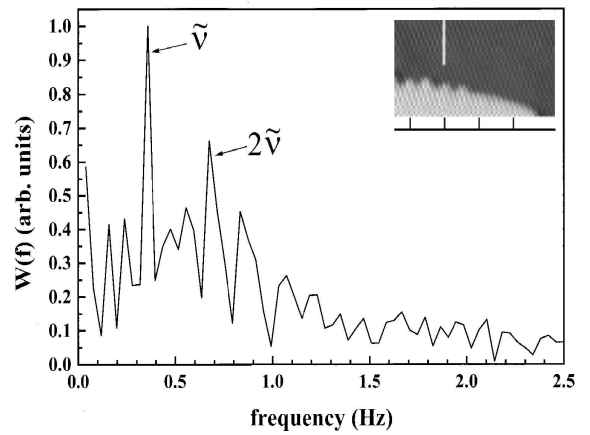}}
\caption{Side-branch formation close to the tip under ``symmetric''
($\xi = 0.5$) square-wave heating as displayed by
a smectic-B dendrite growing into the undercooled nematic phase.
The growth pattern (insert) and the Fourier-spectrum $W(f)$
(main frame) are shown. Black lines in the insert denote 
the forcing frequency $\tilde \nu$. 
The white line marks the position ($53.2 \mu m$ behind the tip), 
where the width $w$ of the dendrite has been measured as a function of 
time. Note the presence of the mode oscillating
with twice the base frequency of forcing; a mode missing from the
forcing spectrum. The relevant parameters are: $\tilde \nu = 0.35$
Hz and $\overline P = 9 \times 10^{-5}$ W/cm$^2$.
}
\label{expdouble}
\end{figure*}
\end{minipage}

\newpage

\begin{minipage}{8.3cm}
\begin{figure*}
\epsfxsize=8cm
\centerline{\epsfbox{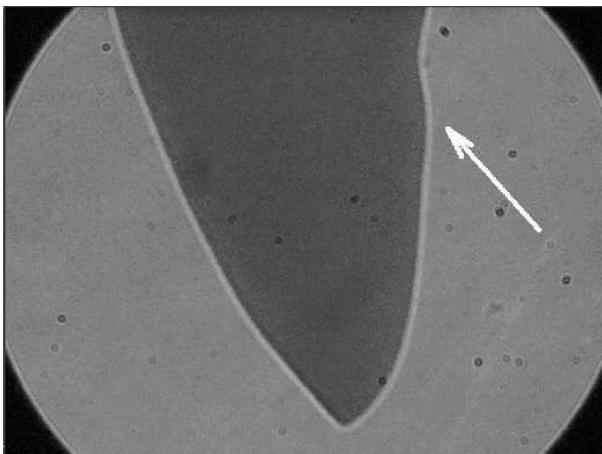}}
\caption{
Shape of the dendrite tip. The arrow points at a surface undulation
that later develops into a side-branch.
}
\label{tipshape}
\end{figure*}
\end{minipage}
%\vspace*{7cm}

%\hspace*{9cm}
\begin{minipage}{8.3cm}
\begin{figure*}
\epsfxsize=8cm
\centerline{\epsfbox{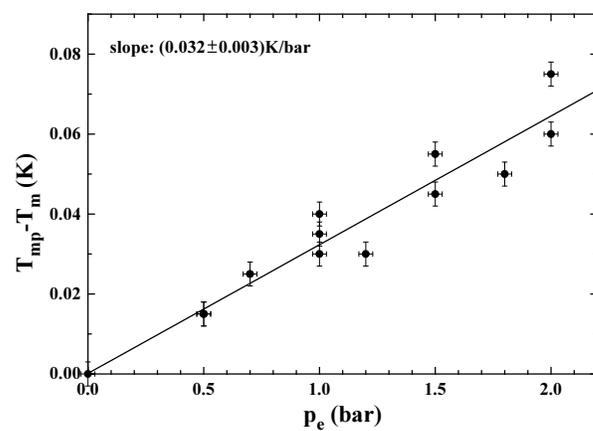}}
\caption{
Pressure dependence of the temperature of the nematic--smectic B phase 
transition.
}
\label{append1}
\end{figure*}
\end{minipage}

\end{multicols}
\end{document}